%% file: paper.tex
\documentclass[10pt,conference]{IEEEtran}
\IEEEoverridecommandlockouts
\usepackage[
style=ieee
]{biblatex}
\usepackage{booktabs}
\usepackage{multirow}
\usepackage{hyperref}
\usepackage{amsmath,amssymb,amsfonts}
\usepackage{algorithmic}
\usepackage{graphicx}
\usepackage{textcomp}
\usepackage{xcolor}
\usepackage{subfig}
\usepackage{comment}
\newcommand{\HA}[1]{{\color{black}{#1}}}
\newcommand{\HH}[1]{{\color{black}{#1}}}

\def\BibTeX{{\rm B\kern-.05em{\sc i\kern-.025em b}\kern-.08em
    T\kern-.1667em\lower.7ex\hbox{E}\kern-.125emX}}

\author{\IEEEauthorblockN{1\textsuperscript{st} Ehsan  Mashhadi}
\IEEEauthorblockA{
\textit{University of Calgary}\\
\textit{Calgary, Canada}\\
\textit{ehsan.mashhadi@ucalgary.ca }
}
\and
\IEEEauthorblockN{2\textsuperscript{rd} Hossein Ahmadvand}
\IEEEauthorblockA{
\textit{University of Calgary}\\
\textit{Calgary, Canada}\\
\textit{hossein.ahmadvand@ucalgary.ca}
}

\and
\IEEEauthorblockN{3\textsuperscript{rd} Hadi Hemmati}
\IEEEauthorblockA{
\textit{York University}
and
\textit{University of Calgary}\\
\textit{Toronto, Canada}\\
\textit{hemmati@yorku.ca}
}
}

\begin{document}

\title{Method-Level Bug Severity Prediction using Source Code Metrics and LLMs}


\maketitle

\begin{abstract}
In the past couple of decades, significant research efforts are devoted to the prediction of software bugs. However, most existing work in this domain treats all bugs the same, which is not the case in practice. It is important for a defect prediction method to estimate the severity of the identified bugs so that the higher-severity ones get immediate attention. In this study, we investigate source code metrics, source code representation using large language models (LLMs), and their combination in predicting bug severity labels of two prominent datasets. We leverage several source metrics at method-level granularity to train eight different machine-learning models. Our results suggest that Decision Tree and Random Forest models outperform other models regarding our several evaluation metrics. We then use the pre-trained CodeBERT LLM to study the source code representations' effectiveness in predicting bug severity. CodeBERT fine-tuning improves the bug severity prediction results significantly in the range of 29\%-140\% for several evaluation metrics, compared to the best classic prediction model on source code metric. Finally, we integrate source code metrics into CodeBERT as an additional input, using our two proposed architectures, which both enhance the CodeBERT model effectiveness \footnote{https://github.com/EhsanMashhadi/ISSRE2023-BugSeverityPrediction, \cite{ehsan_mashhadi_2023_8267597}}.
\end{abstract}

\begin{IEEEkeywords}
Bug severity prediction,
Code representation,
Large Language Models (LLMs),
CodeBERT,
Code metrics.
\end{IEEEkeywords}

\input{conference/introduction}

\input{conference/background}

\input{conference/methodology}

\input{conference/experiments}

\input{conference/related_work}

\input{conference/conclusion}

\section*{References}
\printbibliography[heading=none]

\end{document}

%% file: conference/introduction.tex
\section{Introduction}

Software maintenance including bug handling is one of the most challenging parts of the software development life cycle~\cite{kafura1987use}. Different steps in bug handling such as bug detection, bug localizing, and bug fixing require significant resources (e.g., technical team, time, etc.), and most of these steps are being done manually or semi-automatically in practice~\cite{borstler2016role,bennett2000software}. Consequently, both researchers and practitioners have been trying to automate these tedious tasks from different application perspectives such as defect prediction, fault-localization, test generation, and program repair~\cite{kondo2020impact,shin2010evaluating,tosun2010practical,zhou2010ability,mashhadi2021applying}.

While there has been much research in handling bugs using different techniques like search-based~\cite{le2011genprog}, pattern-based~\cite{long2016automatic}, and ML-based techniques~\cite{mashhadi2021applying}, most of them implicitly assume that all bugs have the same importance which is not true in practice~\cite{shamshiri2015automatically,mashhadi2021applying,wong2016survey,pearson2017evaluating}. Bug severity is one of the bug features that are assigned manually by the QA/developers during the bug report or triaging stages. Bug severity indicates the intensity of bug impacts on the system operation~\cite{neysiani2020efficient}, so it guides technical teams to adjust their priorities regarding how fast the bugs should be fixed~\cite{yang2014towards}. 


Estimating the severity of a given bug, automatically and systematically (not biased to the opinion of the bug reporter), can be quite helpful, not only for the technical teams (including developers and QA teams) but also as an important input for other automated software engineering tools, such as program repair, to prioritize their process, accordingly. 

\HH{
\textbf{Motivation:}
To the best of our knowledge, the current literature in the bug severity prediction domain is mainly limited to studies based on the bug report description using different NLP (Natural Language Processing) techniques~\cite{chaturvedi2012determining,tan2020bug}. These techniques have some limitations, as follows: (a) given that the predicted severities are solely estimated based on the reported severities for a new bug, they may be biased toward the subjective opinion of the reporter (not always an expert); (b) They ignore the context (underlying source code). That is, a similar bug report for different source code might have a different severity; (c) These approaches are only limited to projects with rich and informative bug reports that are not applicable in many cases; and (d) Not all detected bugs have a bug report written for them, even in projects with rich history of bugs. A lot of times, a new bug may be detected by a test case and do not have a report assigned to it to be used for severity prediction. The report will be added eventually after the bug is fixed or assigned to be fixed but the severity is needed before that.

The above challenges motivate us to propose a technique for bug severity prediction that does not require bug reports. This is in fact a harder problem to solve compared to bug report-based alternatives, since (a) the typical tools in hand for prediction (such as transformers and LLMs) are better fits when working with natural language vs source code and (b) when a tester/developer/or even a user writes the bug report they already provide a lot of extra information about the bug including information that may directly or indirectly imply the severity level. Therefore, an approach that does not assume the existence of bug reports is tackling a harder problem. 
}

There is also some limited work on using source code metrics for bug severity prediction ~\cite{zhou2006empirical,shatnawi2008effectiveness,singh2013empirical} which are \HH{only based on class-level source code metrics. However, researchers/developers find the class/module level granularity too coarse-grained for practical usage \cite{shihab2012industrial,pascarella2020performance,grund2021codeshovel,hata2012bug} (class-level results, e.g., which class has more risky bugs, are usually known in advance for developers). In addition, these techniques' estimators are not the state of the art these days (techniques from at least from 10 years ago).}


To overcome the mentioned limitations and advance state of the art, in this paper, we provide an automated approach that leverages the potential of 1) method-level source code metrics, 2) source code representation (code embedding) provided by large pre-trained language models (LLMs) for code, and 3) integration of source code representation and source code metrics for predicting the bug severity values. Our replication package is provided in the online repository \cite{repo} .

We leverage source code representation learning (also known as feature learning) since they have shown significant improvement over the classical models by automatically discovering the features/representations of samples for performing the requested downstream tasks. In recent studies, LLMs for code such as CodeBERT \cite{feng2020codebert} (a large language model based on transformers) have shown promising results on several software-related tasks including bug detection, program repair, and clone detection \cite{lu2021codexglue,mashhadi2021applying}.


\newcommand{\RQFirst}{Are method-level source code metrics able to estimate bug severity?}

\newcommand{\RQSecond}{How well can CodeBERT predict bug severity?}

\newcommand{\RQThird}{Does providing source-code metrics as extra inputs to CodeBERT improve its bug severity prediction?}


\HH{
\textbf{Contribution and Novelty:}}
In this study, we leverage 19 open-source projects containing 3,342 buggy methods. We have trained eight classic machine learning models (KNN, SVM, Naive Bayes, Decision Tree, Random Forest, Ada Boost, XGBoost, and MLP) on the extracted source code metrics and evaluated the results on several performance metrics. The results show that Random Forest has the best performance. Our experiment results on the CodeBERT model, using the source code representation, show significant improvement (29\%-140\% across different metrics) over the previous methods. Finally, we propose to integrate source code metrics and source code representation by providing two architectures, where source code metrics are provided to the model as (a) a textual additional input, and (b) a numerical vector. This approach showed an improvement of (2\%-10\% across different metrics) over the CodeBERT model's effectiveness.
\HH{
The main novelty of this work is on using LLM for bug severity predition without a need for bug reports, which comes with our proposed templates on how to feed source code and method-level metrics together to LLM to get the best results. 

\noindent In summary, the main contributions of this study include:

\begin{itemize}
  \item Investigating eight classic machine learning models trained on 10 method-level source code metrics to predict the bug severity of two popular datasets, as baselines.
  \item Exploring an LLM for code (CodeBERT) in predicting bug severity labels using only buggy method source code.
  \item Proposing a novel approach (two different architectures) for integrating source code metrics with source code representation provided by CodeBERT.
  
\end{itemize}
}

%% file: conference/background.tex
\section{Background}

\subsection{Source Code Metrics}
\label{sec:code_metrics}

Source code metrics have been used for various software-related applications such as code smell detection~\cite{tufano2015and}, maintenance effort estimation~\cite{polo2001using} and defects detection~\cite{ferenc2020deep}. Different granularity levels such as package/class level~\cite{Okutan:2014,Koru:2005,zimmermann2007predicting}, method level~\cite{pascarella2020performance,giger2012method,ferenc:2020,chowdhury:2022,grund2021codeshovel,Mo:2022}, and line level~\cite{wattanakriengkrai2020predicting} have been applied in previous research. High granularity levels (package/class) are practically less helpful for the developers~\cite{giger2012method,pascarella2020performance} since it requires significant effort to locate bugs at the package/class components. In addition, line-level granularity can suffer from too many false positives, because multiple lines can be similar just by chance~\cite{Daniela:2014,Francisco:2017}. Consequently, method level granularity has been the new focus to the community~\cite{pascarella2020performance,giger2012method,ferenc:2020,chowdhury:2022,grund2021codeshovel,Mo:2022}, especially for bug prediction models, and several studies show positive and encouraging results~\cite{giger2012method,ferenc:2020,Mo:2022}.

Due to the mentioned problem, in this study, we use the method-level granularity which is described in this section. These code metrics have different advantages (e.g., being fast to calculate) and disadvantages (e.g., requiring a language-specific parser). 
We made sure we have most of the metrics that have been shown effective in predicting the method-level bugginess, in the previous studies. These metrics are defined as follows:

\textbf{Lines of Code (LC)} also known as Size is the most popular, easy to measure, and the most effective code metric for estimating software maintenance~\cite{Gil:2017,Emam:2001,chowdhury:2022}. In this study, we calculate LC as the source lines of code without comments and blank lines, similar to~\cite{landman2014empirical,ralph2018construct,chowdhury:2022} to prevent the code formatting and comments effects, which are out of this study's scope.

\textbf{McCabe (MA)} also known as cyclomatic complexity \cite{mccabe1976complexity,landman2014empirical}, is another very popular metric that indicates the number of independent paths, and thus the logical complexity of a program. Intuitively, components with high McCabe values are more bug-prone.

\textbf{McClure (ML)} was proposed as an improvement over McCabe \cite{mcclure1978model,kafura1987use}. Unlike McCabe, McClure considers the number of control variables, and the number of comparisons in a predicate, which is not supported by McCabe. 

McCabe and McClure do not consider nested depth. \textbf{Nested Block Depth (NBD)}~\cite{kasto2013measuring,alenezi2019impact,zaw2020software} has been studied alongside them to alleviate this issue.

Since McCabe-like complexity measures require a language-specific parser (for finding the predicates), Hindle et al. \cite{hindle2008reading} proposed \textbf{Proxy Indentation (PI)} as a proxy for McCabe-like complexity metrics.

\textbf{FanOut (FO)} calculates the total number of methods called a given method. This provides an estimate of the coupling, i.e., dependency of a particular method on other methods.

\textbf{Readability (R)} combines different code features to calculate a single value for estimating code readability. We used the readability metric proposed by Buse et al.~\cite{buse2009learning} which generates a readability score for a given method. The readability scores range from 0 to 1 for specifying least readable code to most readable code, respectively.

\textbf{Halstead Metrics} contain seven measures based on the number of operators and operands in a component \cite{halstead1977elements}. As all the Halstead metrics are highly correlated to each other, in this study, we consider only two of them: \textbf{Difficulty (D)} and \textbf{Effort (E)} which use other Halstead metrics in their formulas, as follows:

\begin{displaymath}
E = D * V   \qquad D = \frac{n1}{2} * \frac{N2}{n2}
\end{displaymath}

\begin{displaymath}
N = N1 + N2        \qquad       n = n1 + n2
\end{displaymath}

\begin{displaymath}
V (Halstead Volume) = N * \log_2 (n)
\end{displaymath}

Where~$n1$ and ~$n2$ are the number of distinct operators and operands, and ~$N1$ and ~$N2$ are the total number of operators and operands, respectively.

\textbf{Maintainability Index (MI)} 
has been introduced by Omran and Hagemeister~\cite{oman1992metrics} where the authors defined metrics for measuring 
the maintainability of a software system and combine those metrics into a single value. MI can be calculated as:
\begin{displaymath}
\begin{split}
MI = 171- 5.2 \times ln(Halstead Volume)-\\ 0.23 \times (McCabe)- 16.2 \times ln(LC)
\end{split}
\end{displaymath}
Where~$HalsteadVolume$ and~$LC$ are defined previously in this section.\\

\subsection{Representation Learning}

CodeBERT~\cite{feng2020codebert} is a pre-trained large language model (LLM) which supports both programming language(PL) and natural language (NL). It follows the BERT~\cite{devlin2018bert} and RoBERTa~\cite{liu2019roberta} models architecture and uses a multi-layer bidirectional transformer~\cite{vaswani2017attention}. CodeBERT has the same model architecture as RoBERTa-base with 125M parameters. It has been trained on 6.4M NL-PL pairs of 6 programming languages such as Python, Java, JavaScript, PHP, Ruby, and Go from the CodeSearchNet dataset~\cite{husain2019codesearchnet}. This model has been trained regarding two objectives, masked language modeling (MLM) and replaced token detection (RTD). The input format follows the concatenation of two segments (first part containing natural language, and the second part containing code) with a separator token: $[CLS], w\textsubscript{1},w\textsubscript{2},…,w\textsubscript{n},[SEP],c\textsubscript{1},c\textsubscript{2},...,c\textsubscript{m},[EOS]$. The output includes contextual vector representation of both NL and PL segments, and $[CLS]$ representation can be used as the aggregated sequence representation for various tasks such as ranking or classification. 

CodeBERT has been used for different software engineering tasks such as \cite{mashhadi2021applying} where authors use the CodeBERT model to generate fixes for simple Java bugs with accuracy in the range of 19-72\%. Pan et al.\cite{pan2021empirical} leveraged CodeBERT for bug detection in cross-version and cross-project settings and they also investigated the effects of different prediction
patterns. Zhang et al. \cite{zhang2022improving} used CodeBERT for enhancing the performance of the question title generation task by using bi-modal information that exists in the question body. They proposed that this method outperforms the previous models under different settings in the range of 4\%-14\%.

%% file: conference/methodology.tex
\section{Study Setup}
\subsection{Dataset Collection}
\label{sec:Dataset}

We use Defects4J~\cite{just2014defects4j} and Bugs.jar~\cite{saha2018bugs} datasets containing real bugs from different popular open-source Java projects. Defects4J is selected due to its diversity by having projects in different domains and also its popularity in different automated software engineering research domains such as Fault Localization \cite{pearson2017evaluating}, Program Repair \cite{martinez2017automatic}, and Test Generation \cite{shamshiri2015automatically}. Bugs.jar is another popular dataset containing large open-source projects which have been used widely in Program Repair~\cite{saha2017elixir}.

The Defects4J latest version (2.0.0) contains 835 bugs from 17 Java open-source projects. After extracting bugs having severity labels, we concluded with 510 bugs from projects: Chart, Cli, Closure, Codec, Collections, Compress, Csv, JxPath, Lang, Math, and Time. The Bugs.jar dataset contains 1,158 bugs from eight large and popular open-source Java projects: Accumulo, Camel, Commons-math, Flink, Jackrabbit-oak, Logging-log4j2, Maven, and Wicket.

Our study requires buggy methods along with their severity labels. Therefore, we gathered the severity labels of each bug from its corresponding issue management systems. The defects4J dataset contains bugs from Jira, Google Code Archive, and Source Forge, but the Bugs.jar dataset contains bugs only from Jira. For extracting the severity labels we extract the issue id of a given bug (each bug in both datasets has a unique issue id), and then fetch its severity labels from the corresponding issue management systems. For extracting severity labels from Jira we use ``Python Jira library'' \cite{jira}, for Source Forge and Google Code Archive we parse the web page containing the issue and extract the severity labels. The code regarding dataset collection is available in the provided replication package.

We unify all the categorical and numerical severity labels into the numerical format range from 0 to 3 (0 most critical, 3 least critical) to use them as the target label in our training purpose according to Table \ref{tab:targetvar_severity}. We also merge two datasets and then shuffle the samples to have a relatively larger number of samples (better for training our models) and also reach one model capable of handling diverse bugs from different projects and different domains rather than having a specific trained model for each dataset/project.

\begin{table}[htbp]
    \caption{Unifying severity labels extracted from issue management systems into four numerical severity levels.
\label{tab:targetvar_severity} }
    \begin{center}
      \begin{tabular}{l|l}
        \toprule
        \textbf{Fetched Severity Labels} & \textbf{Class Labels}\\
        \midrule
        Critical, Blocker & 0\\
        Major, High & 1\\
        Medium & 2\\
        Low, Trivial, Minor & 3\\
        \bottomrule
      \end{tabular}
    \end{center}
  \end{table}

Each buggy project in the mentioned datasets exhibits one bug which is fixed in its corresponding fix version. Since our study focus is the method-level granularity, we consider all methods that are impacted during the bug fixing patch as the buggy method, similar to earlier studies (e.g.,~\cite{chowdhury:2022,pascarella2020performance,Mo:2022}). Finally, after merging these two datasets and removing the duplicate instances (the same buggy methods that appear in a few projects/datasets) we reach 3,342 buggy methods. For the training process, we split our dataset randomly in three different train, validation, and test splits using 70\%, 15\%, and 15\% ratios respectively. Note that in our case the random splitting is not considered a data leakage, since bug severities are not correlated over time (knowing a future severity label will not give an extra hint to predict a current severity label vs knowing a past severity label). However, in practice one could only use past data for training (we could not implement the experiments this way since not all our data items had time stamps). Our dataset statistics including the number of instances in each project, train, validation, and test splits regarding the severity labels are provided in Table \ref{tab:dataset_statistics}.

\begin{table}[htbp]
\caption{
\label{tab:dataset_statistics} Dataset statistics including the number of samples with different severity labels in each project, the whole datasets, train, validation, and test splits.
}
      \begin{center}
      \begin{tabular}{l|c|c|c|c}
        \toprule
        \textbf{Project} & \textbf{Class 0} & \textbf{Class 1} & \textbf{Class 2} & \textbf{Class 3}
        \\
        \midrule
        Chart & 0 & 0 & 5 & 3 \\
        Cli & 2 & 36 & 0 & 12 \\
        Closure & 1 & 28 & 269 & 5\\
        Codec & 1 & 16 & 0 & 3\\
        Collections & 0 & 4 & 0 & 0\\
        Compress & 2 & 47 & 0 & 15 \\
        CSV & 0 & 11 & 0 & 3 \\
        JxPath & 1 & 35 & 0 & 4\\
        Lang & 7 & 56 & 0 & 23\\
        Math & 11 & 87 & 0 & 34 \\
        Time & 0 & 1 & 17 & 3\\
        Accumulo & 54 & 146 & 0 & 115\\
        Camel & 8 & 183 & 0 & 111\\
        Commons-math & 24 & 233 & 0 & 66\\
        Flink & 61 & 82 & 0 & 24\\
        Jackrabbit-oak & 82 & 518 & 0 & 96\\
        Logging-log4j & 5 & 84 & 0 & 19 \\
        Maven & 7 & 56 & 0 & 15 \\
        Wicket & 9 & 459 & 0 & 143\\
        \midrule
        All & 275 & 2082 & 291 & 694 \\
        \midrule
        Train & 198 & 1524 & 203 & 489\\
        Validation & 30 & 264 & 40 & 92 \\
        Test & 47 & 294 & 48 & 113\\

        \bottomrule
      \end{tabular}
    \end{center}
    \vspace{-0.5cm}
  \end{table}

\subsection{Source Code Metrics Calculation}
We found that many of our buggy methods in the datasets contain code comments. Since studying code comments' impact on our studied subject is out of the scope of this study, we remove any comment (such as JavaDoc or inline format) that exists in all of our samples.

Furthermore, after computing all of the source code metrics, to remove the potential scaling problem of the numerical data, we use the $Robust Scaler$\cite{robustscaler}. We did this pre-processing step because our calculated source code metric values do not have the same range, so we normalize them by using the $Robust Scaler$ algorithm, which normalizes the input data and is also robust against outliers. This algorithm follows a similar algorithm to the $MinMax$ scale, but it uses the interquartile range rather than the min-max \cite{robustscaler}. 



\subsection{Evaluation Metrics}

Since our dataset is imbalanced (the number of bugs in each severity category is not equal), like many real datasets, we cannot rely on the accuracy metric only. Therefore, we evaluate our experiment using several metrics that are robust against an imbalanced dataset as follows: 

\HH{The first metrics to report are Precision and Recall. Since we have multi-class classifiers (our target classes are 0,1,2,3 that are provided in Table \ref{tab:targetvar_severity} so the provided evaluation metrics such as precision, recall, and ROC-AUC are calculated using the ``weighted" version instead of the default ``binary". This approach calculates metrics for each label and then finds their average weighted by support (the number of true instances for each label) that takes label imbalance into account.





\textbf{F1-Weighted}: 
 F-measure or F-score is a popular evaluation metric in classification tasks that considers both precision and recall. F1 is the most common form of F-measure, which takes the harmonic mean of precision and recall.
 However, we cannot use the F1 score directly for multi-class problems, and since our dataset is also imbalanced, the F1-weighted metric is used. It is calculated by taking the mean of all per-class F1 scores and considering each class sample number.

}

\textbf{ROC Curve Plot}:
The Receiver Operator Characteristic (ROC) is a curve that plots the True Positive Rate (TPR) against the False Positive Rate (FPR) at various threshold values. FPR is defined as FP divided by FP+TN. 


Using this plot, a perfect model is represented by a line from the bottom left to the top left and then across the top right of the plot. A skillful model is represented by a curve that bows up to the top left of the plot. However, a random classifier (no skill) is represented by a diagonal line from the bottom left to the top right of the plot.

\textbf{AUC-ROC}: The Area Under the Curve (AUC) is the measure of the ability of a classifier to distinguish between target classes. The perfect classifier has an AUC score of 1 which means it distinguishes between the positive and negative classes well while the classifier with an AUC score of 0.5 means it works like a random classifier. AUC score of 0 means that the classifier predictions are 100\% wrong.

\textbf{Precision-Recall Curve Plot}: The precision-Recall curve shows the trade-off between precision and recalls for different thresholds. The point in the top right shows a perfect classifier while a no-skill(random) classifier will be a horizontal line on the plot with a precision that is proportional to the number of positive examples \cite{classification_metrics}.

\textbf{MCC}: Matthews Correlation Coefficient (MCC)~\cite{matthews1975comparison} assess the performance of a classifier by leveraging all of the values available in the confusion matrix that makes it helpful for imbalanced datasets. It is calculated as below:

\begin{displaymath}
\frac{TP*TN - FP*FN}{\sqrt{(TP+FP)*(TP+FN)*(TN+FP)*(TN+FN)}}
\end{displaymath}

The MCC value is between -1 and +1 where +1 represents a perfect prediction, 0 represents a random (no-skill) prediction, and -1 represents an inverse prediction.

\HH{Note that all evaluation metrics including the weighted versions are implemented using Scikit-learn's standard metrics library, where the exact formula per metric can also be found in its documentation \cite{SKLearnEval} }

%% file: conference/experiments.tex
\section{Experiments}

\subsection{RQ1: \RQFirst\ }

\textbf{Objective:} In this RQ, we aim to study our selected source code metrics, which have shown significant results in defect detection tasks in previous studies, and their ability in predicting bug severity using several machine learning models. To the best of our knowledge, there is no previous study on predicting bug severity labels on Defects4J and Bugs.jar datasets (relatively small datasets of open-source real-world projects) using method-level source code metrics. Our motivation behind this RQ is assessing machine learning models' capability on the set of diverse projects using method-level granularity (since it is more efficient for practitioners to handle bugs at this level compared to the class/package granularity and the line-level defect prediction is not mature-enough yet).

\textbf{Design:} We train eight machine learning models such as KNN, SVM, naive Bayes, Decision Tree, RandomForest, XgBoost, AdaBoost, and MLP by using our source code metrics which are described in section \ref{sec:code_metrics}. By choosing these eight models, we cover a diverse set of classification algorithms that have shown promising results in research/practice for various classification tasks. Since the hyper-parameter tuning is out of the scope of this study we keep the default hyper-parameters (available in the Scikit-learn library implementation \cite{buitinck2013api}) for all of these models.

Note that we consider RQ1 as our baseline of comparison for the next two RQs. Even though applying the exact same models on the exact same datasets, are not dine before and is our contribution, but such classic methods have been widely used on defect prediction and in limited cases on bug severity prediction in class level, in the past. Thus it is fair to consider them as our baseline of comparison.

\textbf{Result:} Table \ref{tab:baseline_metrics} shows the results of our studied models regarding different evaluation metrics (due to the limited space in this paper, we provide the confusion matrix, ROC, and Precision-Recall curves of these eight models in our provided code repository). In this study, we only report F1 for each class (F1 Perclass) that shows the models effectiveness for each target class (different severity groups) separately, but provide the ``weighted" versions of Precision, Recall, AUC, and MCC since the ``weighted" approach takes all of the target classes (different severity groups) into consideration and provides a better estimation of the model effectiveness in predicting different severity groups. Since all models have an AUC value $>$ 0.5, we can say that they work better than a random classifier. However, as is shown in the table (highlighted with a bold style), the Random Forest model has the best effectiveness in predicting bug severity labels by having better results using all provided evaluation metrics. In contrast, the SVM model has one of the weakest performances in predicting the severity values, based on any metric. According to its MCC value, this model works like the no-skill (random) classifier (at least with the default parameters of the Scikit-learn library). The F1 per-class metric shows that the SVM model can only predict the ``class 1 (Major, High) severity" correctly and miss the other severity groups. Furthermore, its low precision value (0.34) shows that it returns a lot of false positives. One possible reason behind the relatively poor results of SVM and Naive Bayes models is their inherent need of having a powerful set of features and enough samples. Since our dataset size is relatively small for training a model from scratch, so replicating this study using other datasets/features may improve the results of these models.

As mentioned, previous studies~\cite{zhou2006empirical,shatnawi2008effectiveness,singh2013empirical} have leveraged some object-oriented source code metrics that are different from our method-level source code metrics. However, we cannot simply compare our RQ1 results to these published works since (1) they are not at the method level, which is the more practical level from practitioners' perspective~\cite{pascarella2020performance,grund2021codeshovel}, and (2) they are predicting on different severity levels (typically only ``high'' and ``low'' levels). \\


\noindent\fbox{%
    \parbox{.98\columnwidth}{%
    \textbf{RQ1 Summary:}
Among the classic models, Random Forest (MCC=0.21) and SVM (MCC=0) models have the best and weakest performance, respectively.
    }%
}\\

\begin{table*}[htbp]
\caption{
\label{tab:baseline_metrics} Trained models' performance with different evaluation metrics. The bold line shows the best value in each column and the Random Forest model shows the best overall performance regarding F1 Weighted, AUC, and MCC metrics.
}
      \begin{center}
      \begin{tabular}{l|l|l|l|l|l|l}
        \toprule
        \textbf{Model} & \textbf{Precision (Weighted)} & \textbf{Recall (Weighted)} & \textbf{F1 (Weighted)} & \textbf{F1 Per-class [0,1,2,3]} &\textbf{AUC (Weighted)}
        &\textbf{MCC}
        \\
        \midrule
        KNN & 0.49 & 0.56 & 0.49 & [0.16, 0.71, 0.15, 0.19] &  0.60 & 0.11 \\
        SVM & 0.34 & 0.58 & 0.43 & [0.00, 0.73, 0.00, 0.00] & 0.55 & 0.00 \\
        Naive Bayes & 0.45 & 0.53 & 0.43 & [0.05, 0.71, 0.05, 0.05] & 0.52 & 0.02\\
        Decision Tree & 0.51 & 0.52 & 0.51 & [\textbf{0.36}, 0.65, \textbf{0.23}, \textbf{0.34}] & 0.59 & 0.17 \\
        \textbf{Random Forest} & \textbf{0.59} & \textbf{0.60} & \textbf{0.54} & [0.35, 0.73, 0.20, 0.26] & \textbf{0.65} & \textbf{0.21}\\
        Ada Boost & 0.43 & 0.56 & 0.44 & [0.00, 0.72, 0.07, 0.06] & 0.56 & 0.01 \\
        XGBoost & 0.52 & 0.57 & 0.51 & [0.29, 0.71, 0.22, 0.20] & 0.63 & 0.14 \\
        MLP & 0.47 & 0.59 & 0.45 & [0.00, \textbf{0.74}, 0.10, 0.04] & 0.59 &  0.10 \\
        \bottomrule
      \end{tabular}
    \end{center}
  \end{table*}

\subsection{RQ2: \RQSecond\ }
\textbf{Objective:} The purpose of this RQ is to study the effectiveness of code representation provided by a large pre-trained language model named CodeBERT in predicting bug severity.

The motivations behind this RQ are: 1) these models have outstanding abilities in wide-range of software-related tasks including bug detection, clone detection, and program repair that are shown successfully in previous studies \cite{lu2021codexglue}, 2) since these models have been trained on millions of samples, it enables us to fine-tune them for various downstream tasks (e.g., our bug prediction tasks) with relatively small datasets (such as our dataset) compared to training models from scratch that requires large datasets, 3) these models leverage representation of input data, so there is no need to extract features from input data (such as extracted code metrics in RQ1), which is a hard task in complicated data and applications, and finally, 4) since these models have been trained on several programming languages, our study can be easily applied to other programming languages (supported by CodeBERT) without any changes. 

Furthermore, CodeBERT supports both programming language and natural language inputs, so additional inputs such as bug descriptions can be provided as additional input for our studied task that may improve the results further, which can be studied as a potential future work.

\textbf{Design:} We select the ``codebert-base'' model from Hugging Face \cite{hugging-face} which is initialized with ``roberta-base'' and trained on bi-modal data (documents \& code) with MLM (masked language modeling) and RTD (replaced token detection) objectives. Since we only use source code as an input, we do not leverage the NL(Natural Language) part of this model's input.

Since the CodeBERT model is an encoder we append a $RobertaClassificationHead$ layer, to make it applicable for our classification task. We add this layer to the CodeBERT to follow the architecture named $RobertaForSequenceClassification$, which is provided by the Hugging Face as a classification model. We use the Softmax layer in this architecture since we have a multi-class classification problem and also the CrossEntropyLoss is used for calculating the loss during the training process.

For fine-tuning the CodeBERT model, we first freeze its layers to prevent changing the weights of the existing layers drastically and only train our added classification layer by a large number of epochs (40 epochs). During the training process, we use the early stopping technique with $patience=3$ to prevent the over-fitting problem by considering the model's loss value using the validation set. In the next step, we unfreeze the CodeBERT model's layers and fine-tune the whole architecture by using a smaller learning rate and fewer epochs to adjust all layers' weight. In this step, we use five epochs, as it is suggested in previous research \cite{devlin2018bert} to use a small number of epochs to prevent the over-fitting problem and not change the pre-trained model weights, drastically. During the whole training process, we use the train split for training the model and the validation split for evaluating the model's effectiveness, and we also save the checkpoint of the model with the best F1-weighted score. Finally, we use the test split (which is never seen during the training process) for the model's final evaluation.

Since 512 block size is the maximum value supported by the CodeBERT, we pad our input with the $padding$ token whenever it is shorter than 512 tokens, and whenever the sample length is larger than 512 we truncate it by keeping the first 512 tokens.

\begin{figure}[]
\centering
  \begin{minipage}{0.85\columnwidth}
     \centering
     \includegraphics[width=1\columnwidth]{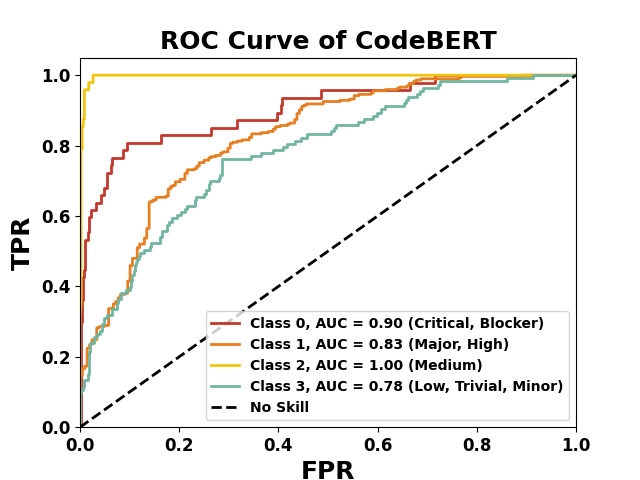}
  \end{minipage}\hfill
  \begin{minipage}{0.85\columnwidth}
     \centering
      \includegraphics[width=1\columnwidth]{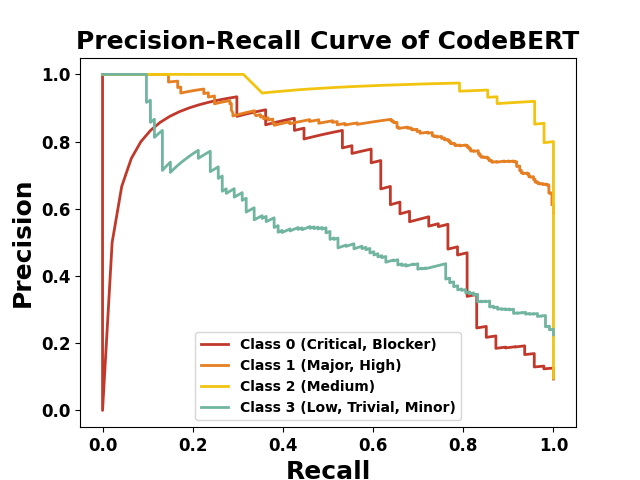}
  \end{minipage}\hfill
\caption{ROC curve and Precision-Recall curve of CodeBERT.}
\label{fig:precision_recall_codebert}
\vspace{-0.5cm}
\end{figure}

\textbf{Result:} The ROC and Precision-Recall curves of the CodeBERT model are provided in Fig \ref{fig:precision_recall_codebert}. These graphs show that the CodeBERT model predicts the ``Class 2 (Medium)" severity label very well since the ROC and Precision-Recall curves of this class is very close to the perfect classifier curve. This result seems interesting since, despite the low number of ``Class2 (Medium)" instances in our dataset (489 in the train split), the model has predicted most of these class instances (42 out of 48) in the test split correctly.

Table \ref{tab:codebert_metrics} shows result of this model regarding our evaluation metrics. These results show CodeBERT's promising effectiveness in predicting bug severity due to relatively high values of all evaluation metrics. By comparing this result to the eight different models' results provided in RQ1(Table \ref{tab:baseline_metrics}), we can state that the CodeBERT model has significantly better effectiveness since it increases the F1-Weighted, AUC, and MCC values by 0.16 (29\%), 0.22 (33\%), and 0.30 (140\%) respectively compared to the Random Forest model (the best model among all RQ1 models).

 \begin{table*}[htbp]
\caption{
\label{tab:codebert_metrics} CodeBERT model performance with different evaluation metrics.
}
      \begin{center}
      \begin{tabular}{l|l|l|l|l|l|l}
        \toprule
        \textbf{Model} & \textbf{Precision} & \textbf{Recall} & \textbf{F1-Weighted} & \textbf{F1-Perclass} &\textbf{AUC}
        &\textbf{MCC}
        \\
        \midrule
        CodeBERT & 0.71 & 0.72 & 0.70 & [0.63, 0.79,   0.90, 0.42] & 0.87 & 0.51\\
        \bottomrule
      \end{tabular}
    \end{center}
    \vspace{-0.5cm}
  \end{table*}

\begin{figure}[]
\centering
\includegraphics[width=0.85\columnwidth]{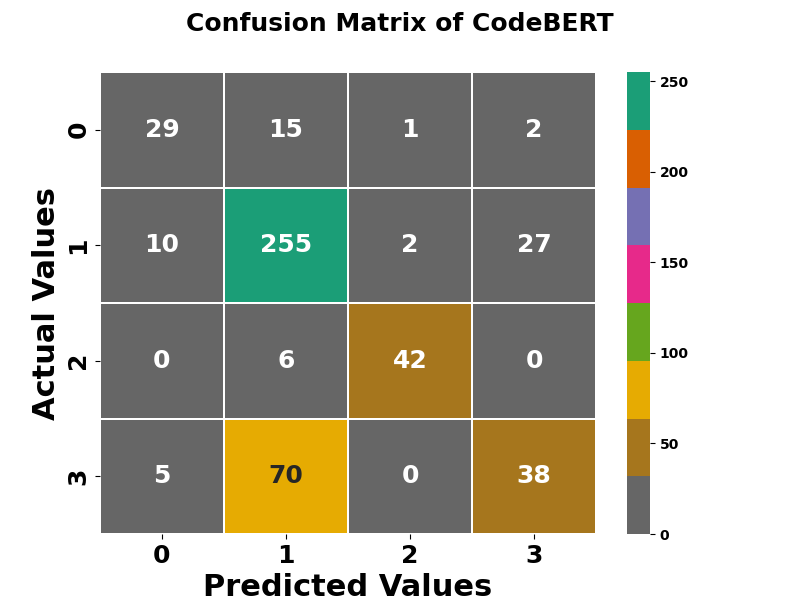}
\caption{Confusion matrix of the CodeBERT model.}
\label{fig:confusion_matrix_codebert}
\vspace{-0.5cm}
\end{figure}

Furthermore, we have provided the confusion matrix of the CodeBERT model in Fig \ref{fig:confusion_matrix_codebert} where the actual labels and the predicted labels are shown for each class. This figure shows that CodeBERT mislabeled 18 bugs with ``class 0" severity; it predicts 15 of them as the ``class 1" severity, which means that it detected critical bugs as major severity bugs, but it assigns lower severity (medium, low) to only three bugs. For the bugs having ``class 1" labels, it predicted the severity in 87\% cases correctly. It also classified all of the bugs with the ``class 2" label correctly, except for six bugs where it assigned higher severity values. The bugs with the ``class 3" labels are predicted to have higher severity values in many cases. These miss-classifications may increase the developers/QA team's investigation time in practice, but it does not increase the risk of using this technique as opposed to situations where the model assigns the lower severity labels to the actual bugs (e.g., considering critical bugs as low-severity bugs). It also still performs better than the best classical method (Random Forest) even in terms of miss-classification for this class.

Our results suggest that code representation has a higher capability in predicting bug severity of studied datasets compared to using our selected source code metrics, solely. The first potential reason is the power of learned semantics behind the source code embeddings and code representation compared to the extracted source code metrics that may not have a high correlation with the bug severity, regardless of the predicting model. The second potential reason is, in cases where the dataset size is relatively small, training a model from scratch may not lead to good results, but using pre-trained models and fine-tuning for the requested downstream task works well since it has already been trained on large datasets during the pre-training phase.\\

\noindent\fbox{%
    \parbox{.98\columnwidth}{%
    \textbf{RQ2 Summary:}
CodeBERT outperforms Random Forest (RQ1's best model) by 29\%, 33\%, and 140\% for F1-Weighted, AUC, and MCC, respectively.
    }%
}\\

\subsection{RQ3: \RQThird\ }
\textbf{Objective:} In this RQ, we study the impact of integrating the source code metrics with the CodeBERT models. Since we found that source code metrics solely are not good indicators of bug severity in our datasets, the motivation behind this RQ is finding their capability as an additional input (since they exhibit important features regarding the source code such as complexity, readability, maintainability, etc.) to the CodeBERT model to improve the model's effectiveness.

\textbf{Design:} \HH{Since CodeBERT has been trained on source code, it can simply understand the source code input while our calculated source code metrics are several numeric values (e.g., 1, 2) so our challenge if to find ways to integrate these numerical values with our source code input in a way that CodeBERT can understand and leverages these data. We provided two different architectures named "ConcatInline" and "ConcatCLS". }

For the ConcatInline architecture, we convert our source code metric features to the textual format and provide it along with the source code input to the model. There are many potential ways of adding numerical/categorical features to the CodeBERT model. Our idea is based on previous research in NLP where numerical features are added to BERT \cite{devlin2018bert} for short text classification~\cite{hu2022short}. Since the CodeBERT model has been trained on the textual data and also considers the context around each word in the training/inference phases, we provide context for each source code feature to make the whole input rich and informative. In this way, instead of providing raw code metrics values directly, we provide a simple paragraph containing a description (based on a template text) for each code metric and also the code metric numerical values. We did not convert the code metrics' values to the letter format (e.g. using ``one'' instead of ``1'') since the letter format increases the input text length drastically and the CodeBERT model has a restriction on the input size.  An example of this approach is shown in Table \ref{tab:context_metrics}. Note that using other templates for the context may change the results, and proper sensitivity analysis to the context template is for our future work.

\begin{table}[htbp]
    \caption{The paragraph contains the context for each source code metric and their values to be used as the natural language(NL) part of the model's input.
\label{tab:context_metrics} }
    \begin{center}
      \begin{tabular}{l|l|l}
        \toprule
        \textbf{Code Metric} & \textbf{Value} & \textbf{NL Part of Model Input}\\
        \midrule
        LC & 1 & \multirow{10}{*}{
        \begin{minipage}{0.25\textwidth} The code contains 1 lines and its complexity metrics values are 2, 3 and 4. The nested block depth is 5,
    and the difficulty of this code is 6. The maintainability score is 7 and this method calls 8 number of methods while its readability and effort metrics values are 9, 10\end{minipage}
    }\\
        PI & 2\\
        MA & 3\\
        NBD& 4\\
        ML & 5\\
        D & 6\\
        MI & 7\\
        FO & 8\\
        R & 9\\
        E & 10\\
        \bottomrule
      \end{tabular}
    \end{center}
  \end{table}

After creating the paragraph describing code metrics and their values we put it in the first section of the CodeBERT model's input (before the $SEP$ token) and then provide the source code containing the buggy method in the second section of the model's input (after $SEP$ token). The final input structure to CodeBERT in this architecture is:
$[CLS],Created Paragraph,[SEP],Source Code,[EOS]$

Since the CodeBERT model can support input up to 512 tokens, in case of long inputs, we keep the whole section of NL (natural language containing the created paragraph of code metrics) and truncate the PL (programming language part containing buggy method source code) section from the end to reduce the whole input size to 512 tokens. In contrast, if the length of the final input is smaller than 512, we pad the input to the end with the $Pad$ tokens.

\begin{figure}[]
\centering
\includegraphics[width=0.5\textwidth]{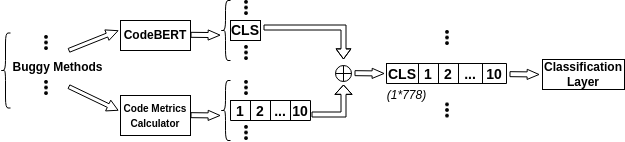}
\caption{ConcatCLS architecture. For each sample, the CLS vector size is (1*768) and the code metric vector size is (1*10).}
\label{fig:concat_cls_architecture}
\vspace{-0.5 cm}
\end{figure}

For the $ConcatCLS$ architecture, we integrate the numerical source code metrics vectors with the CodeBERT model's output $CLS$ token which is shown in Fig \ref{fig:concat_cls_architecture}. In other words, in this architecture, we concatenate the CodeBERT model's output ($CLS$ token exhibiting source code representation) and our numerical vectors (representing source code metrics) and feed the result vector to a dense layer. Having this architecture, during the back-propagation phase of the fine-tuning process, the CodeBERT model weights are not only updated according to the source code representations but they are also updated based on the source code metrics contributions.

\begin{figure*}
  \begin{minipage}{0.5\textwidth}
     \centering
     \includegraphics[width=0.8\textwidth]{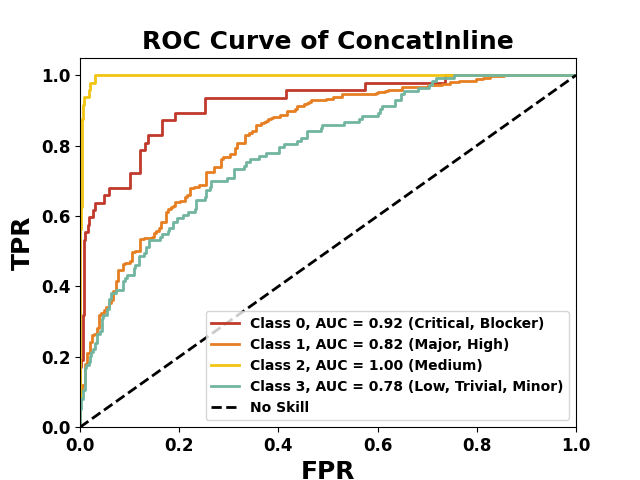}
  \end{minipage}
  \begin{minipage}{0.5\textwidth}
     \centering
      \includegraphics[width=0.8\textwidth]{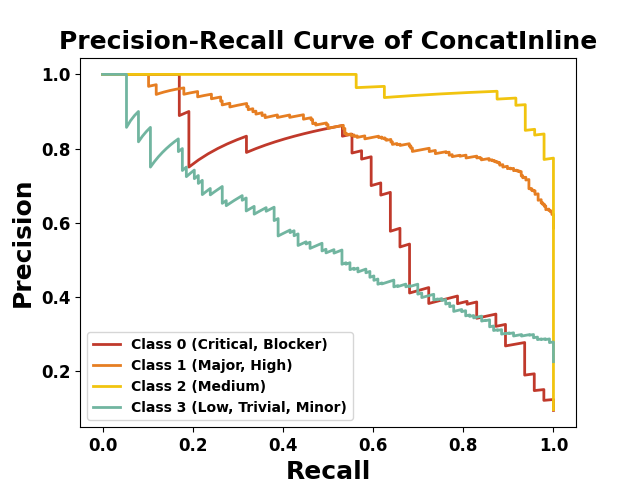}
  \end{minipage}
  \begin{minipage}{0.5\textwidth}
     \centering
     \includegraphics[width=0.8\textwidth]{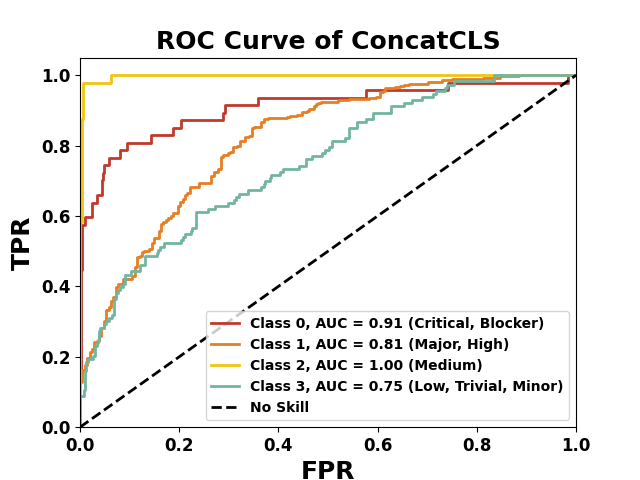}
  \end{minipage}
  \begin{minipage}{0.5\textwidth}
     \centering
      \includegraphics[width=0.8\textwidth]{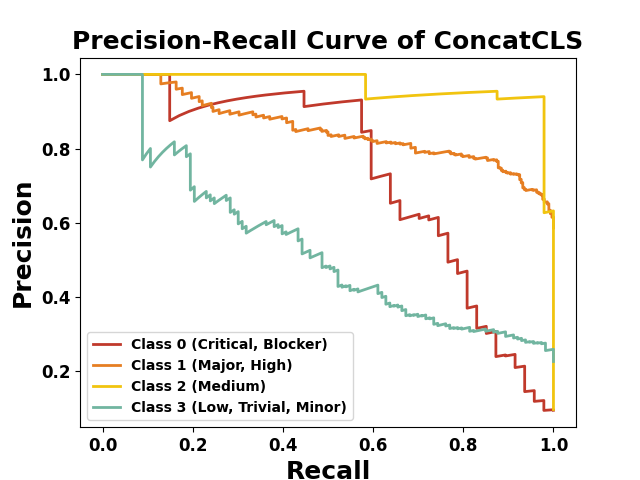}
  \end{minipage}
\caption{ROC and Precision-Recall curves of ConcatInline and ConcatCLS architectures.}
\label{fig:precision_recall_concatinline}
\end{figure*}

\textbf{Result:} The ROC and Precision-Recall curves of our experiments using the $ConcatInline$ and $ConcatCLS$ architectures are shown in Fig \ref{fig:precision_recall_concatinline}. Also, the results regarding the other evaluation metrics are provided in Table \ref{tab:concatinline_metrics}. Our results show that integrating source code metrics in the textual format ($ConcatInline$) increases the effectiveness in predicting different severity labels of the CodeBERT model by 0.03, 0.01, and 0.04 for F1-Weighted, AUC, and MCC metrics, respectively. In addition, the provided ROC curves of $ConcatInline$ architecture shows that adding source code metrics increases the capability of the CodeBERT model to predict the ``class 0 (Critical, Blocker)" severity label by 0.02 in comparison to the RQ2 result. This improvement shows the decisive impact of our source code metrics in predicting bug severity values, especially for high-severity bugs.

 \begin{table*}[htbp]
\caption{
\label{tab:concatinline_metrics} ConcatInline and ConcatCLS architectures performance regarding different evaluation metrics.
}
      \begin{center}
      \begin{tabular}{l|l|l|l|l|l|l}
        \toprule
        \textbf{Model} & \textbf{Precision} & \textbf{Recall} & \textbf{F1-Weighted} & \textbf{F1-Perclass} &\textbf{AUC}
        &\textbf{MCC}
        \\
        \midrule
        ConcatInline & 0.73 & 0.75 & 0.73 & [0.64, 0.81, 0.91, 0.44] & 0.88 & 0.55\\
        ConcatCLS & 0.73 & 0.75 & 0.74 & [0.68, 0.82, 0.95, 0.47] & 0.87 & 0.56 \\
        \bottomrule
      \end{tabular}
    \end{center}
  \end{table*}

\begin{figure}
\centering
  \begin{minipage}{0.7\columnwidth}
     \centering
     \includegraphics[width=0.8\columnwidth]{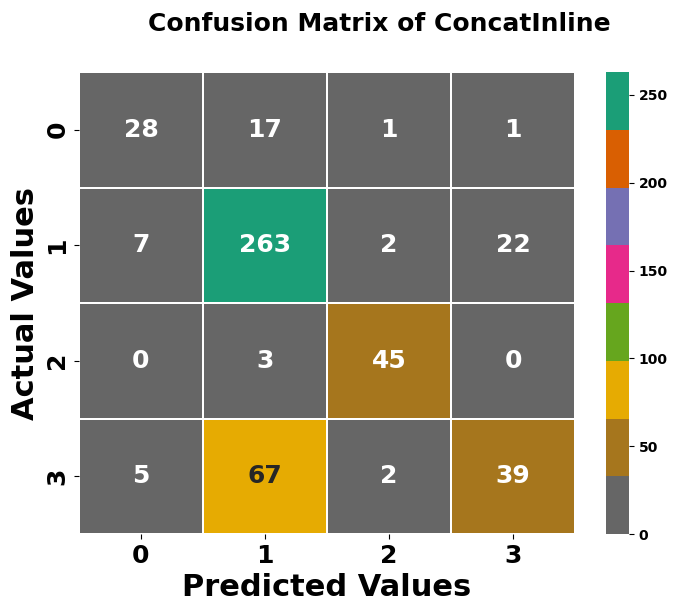}
  \end{minipage}\hfil
  \begin{minipage}{0.7\columnwidth}
     \centering
      \includegraphics[width=0.8\columnwidth]{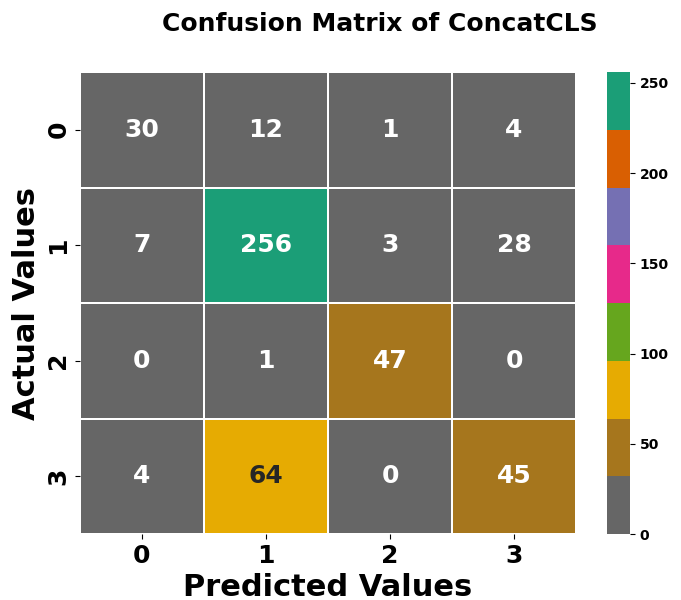}
  \end{minipage}\hfill
\caption{Confusion matrix of ConcatInline and ConcatCLS.}
\label{fig:confusion_matrix_concatinline_concatcls}
\end{figure}

$ConcatCLS$ architecture has increased the effectiveness of the CodeBERT model by 0.04, and 0.05 for F1-Weighted and MCC metrics respectively, but the AUC value is not improved. Also, from the F1-Perclass metric value, we find that the results are improved in all classes. Fig \ref{fig:confusion_matrix_concatinline_concatcls} shows the confusion matrix of both $ConcatInline$ and $ConcatCLS$ architectures. By comparing these figures with the confusion matrix of the CodeBERT model (Fig \ref{tab:codebert_metrics}) we find that the $ConcatInline$ model works better than the CodeBERT model for all classes except for the ``class 0" label where the number of correct classification is reduced by one. The $ConcatCLS$ model has improved the number of correct classifications in all classes (by 1, 1, 5, 7, for class 0, to 3, respectively) compared to the CodeBERT model.

The interesting result from the confusion matrices is that the $ConcatCLS$ architecture has improved the prediction effectiveness for ``class 3 (Low, Trivial, Minor)" since it has decreased the number of miss-classification of this class to ``class 2 (Medium)" (this issue was discussed in the RQ2) from 70 to 64 (8.5\%).

For comparing the $ConcatCLS$ and $ConcatInline$ architectures we consider the confusion matrix and evaluation metrics values. We consider the confusion matrix since it provides better information about each label separately and also other evaluation metrics such as MCC, F1-perclass, and F1-weighted are good metrics for classification tasks, especially the MCC value which works well on imbalanced datasets, like our dataset. Based on the provided figures and tables showing the results, we can state that the $ConcatCLS$ architecture outperforms the $ConcatInline$ model. Finally, we conclude that integrating source code metrics as an additional input in $ConcatInline$ and $ConcatCLS$ architectures improves the CodeBERT model's effectiveness.\\

\noindent\fbox{%
    \parbox{.98\columnwidth}{%
    \textbf{RQ3 Summary:}
Integrating code metrics with the code representation, in a vector format ($ConcatCLS$) improves CodeBERT's effectiveness (MCC) by 5\% and decreases the miss-classification rate by 8.5\%.
    }%
}\\
%
\HH{
\subsection{Threats to Validity}
\textbf{Conclusion validity:} To deal with potential randomness of the results we first removed potential duplicated samples from our dataset, and then shuffle and split it into three train, validation, and test splits using 70\%,15\%,15\% ratio. We did not perform cross-validation technique to keep
our study aligned with previous studied research such as CodeXGLUE and the CodeBERT provided experiments. However, running cross-validation techniques such as K-Fold may affect our obtained results which can be investigated in future work.

\textbf{Internal validity:} One of the internal validity threats to this study is the way we have to deal with the limitations of tools such as context length in LLMs. CodeBert's 512 token size was dealt with by truncating large samples to the first 512 tokens. This may impact the reported results. Future work may try LLMs with larger context size to understand the impact of this limitation.  
Another issue would be the process of integrating numerical features into the CodeBERT model which may affect our results, so for not changing the baseline CodeBERT model structure drastically, we did not add many complex layers to make the comparison fair. However, it is possible to leverage numerical features and integrate them with the CodeBERT model in a more complex way which may improve the results.
We also extracted well-known source code metrics using standards libraries, to reduce any confounding factor in these measurements.

\textbf{Construct validity:} To reduce the impact of construct validity threats we have used seven evaluation metrics. 
The skewed class distribution inherent in imbalanced datasets can result in models exhibiting prediction bias towards the majority classes, thereby evaluating these models is a challenging task. So we use the ``Weighted'' format of our evaluation metrics to reduce the biased effects. These metrics are calculated by taking the mean of all per-class scores and also considering each class sample number. In other words, the calculated output has accounted for each class contribution by considering the number of instances in that given class. 
We also reported all the raw data as confusion matrices.

\textbf{External validity:} Although we have conducted a large experiment with several projects in two distinct datasets, all our projects are from open source domain and future work is needed to replicate the study on commercial applications. In addition, even though at the time of designing this study CodeBERT was the state of the art LLM for code, the community has moved a long way until now then by introducing more advance and larger models such as GPT-4. Future works are needed to compare the results when even more advanced LLMs are hired. However, we only expect better or at least equal results when using more advanced LLMs.

}

%% file: conference/related_work.tex
\section{Related Work}

\textbf{Bug report based} approaches leverage the bug report description and natural language techniques to predict the bug severity. Tian et al.~\cite{tian2012information} use the nearest neighbors classifier and measure the similarity between previous bug reports and the given bug reports to predict the given bug severity label. Their extensive experiments on thousands of bug reports show promising results by having up to 76.3\% for the F-measure metric. Emotion analysis is  another technique leveraged by Ramay et al~\cite{ramay2019deep} where the authors use a deep learning model to analyze the given bug reports for predicting bug severity. Their proposed models have F-measures up to 88.01\%. Lamkanfi et al~\cite{zhang2016towards} proposed the bug severity prediction using the Topic Modeling technique. Their evaluation of 30,000 bug reports extracted from Eclipse, Mozilla, and Netbeans projects shows the effectiveness of their approach by reporting the F-measure of 65\% to 75\% on Bugzilla reports. Tan et al \cite{tan2020bug} leveraged question-and-answer pairs available on the Stack Overflow website and combine them with related bug reports to make an enhanced version of the bug reports. This combination makes the bug report rich which led to improvements by approximately 23\% of the average F-measure metric.

\HA{The most recent work on bug report-based severity prediction \cite{izadi2022predicting}, in fact, uses Transformer models similar to our CodeBERT. However, their research is different from our work for the following reasons: a) they are still based on bug reports only and not source code, and (b) they have a binary classification on ``priority labels'' which is a different field than ``severity'' in issue tracking systems. Although there might be correlations, but severity usually is inherent to the bug (thus to a degree can be estimated by code only) but ``priority'' is assigned partly based on the project’s schedule. In other words, one may tag a bug that is not severe as high priority because e.g., according to the developers schedule it is better to be fixed now than next month, thus having higher priority. 

As explained in the introduction section, unlike all the papers in this category, we do not work on the issue reports but only on source code. Since the assumption that all bugs in the code have an issue report assigned to them is wrong and there are many bugs detected by running test cases that do not have a bug report (or at least not yet).
For the same reason, we did not compare such studies with our approach. }

\textbf{Source code metrics based} approaches extract code metrics from the buggy code and predict the severity labels based on them. Object-oriented class-level design metrics are used for predicting high and low Severity faults on NASA dataset \cite{zhou2006empirical} where the authors find these design metrics are able to predict low severity faults in fault-prone classes better than high severity faults. In another study, various object-oriented class-level metrics are used for training logistic regression and neural network techniques to predict bug severity of three versions of Mozilla Firefox \cite{singh2013empirical}. The authors state that their proposed neural network predicts high and medium-severity errors more accurately than low-severity errors. Several object-oriented class-level metrics are used to investigate the class error proneness in a system’s post-release evolution by using three releases of the Eclipse project \cite{shatnawi2008effectiveness}.

Unlike most of the previous work, we focused on ``method-level'' bug severity prediction which in the real world is much more practical since it can be computed efficiently and also class/module level granularity is too coarse-grained for practitioners to act upon \cite{shihab2012industrial,pascarella2020performance,grund2021codeshovel,hata2012bug}. We also bring in code representation as a state-of-the-art approach to maximize the prediction power from code and metrics.


%% file: conference/conclusion.tex
\section{Conclusion and Future Work}
\HH{In this paper, we propose an approach to use LLM for code to predict the severity of a bug only using the buggy method as input. It integrates extracted method-level source code metrics and the buggy method's code as input and uses CodeBERT to predict multi-class severity labels. Through a large scale empirical study on 3,342 buggy methods from 19 projects of Java open-source projects, we show that using codeBERT only on source code can improve prediction using classic methods on code metrics up to 140\%, for different evaluation metrics. 
In addition, we showed that providing the collected source code metrics  as an additional input to CodeBERT can further improve the results. Interestingly, it also reduces the false positive rate up to 8.5\% for different severity levels, which shows that the pre-trained models must be guided (e.g., by giving the extra information/context and fine tuning) for a specific task to provide the best results.} 
Potential future directions of this research are adding available code comments and inherent structure of code (e.g., data flow, control flow, etc.) as an additional input into the CodeBERT model (or more recent LLMs such as GPT-4) that may enhance the results.\\